\documentstyle[12pt]{article}
\def\baselinestretch{1.5}
\begin{document}
\pagestyle{empty}
%
\vskip 1cm
\centerline{\bf \large
A scheme for radiative CP violation
}\vskip 1cm
\centerline{
Darwin Chang$^{(1,2,3,4)}$ and
Wai-Yee Keung$^{(2,3)}$ }\vskip 1cm
\centerline{\small\it
$^{(1)}$Physics Department,
National Tsing-Hua University, Hsinchu, Taiwan}
\centerline{\small\it
$^{(2)}$Department of Physics, University of Illinois at Chicago,
Illinois 60607--7059}
\centerline{\small\it
$^{(3)}$High Energy Physics Division, Argonne National Laboratory,
Illinois 60439--4815}
\centerline{\small\it
$^{(4)}$Institute of Physics, Academia Sinica, Taipei, Taiwan}
%
\vfill
\begin{abstract}
We present a simple model in which CP symmetry is spontaneously broken
only after the radiative corrections are taken into account. The model
includes two Higgs-boson doublets and two right-handed singlet neutrinos
which induce the necessary non-hermitian interaction. To evade
the Georgi-Pais theorem, some fine-tuning of coupling constants is
necessary.  However, we show that such fine-tuning is natural
in the technical sense as it is protected by symmetry.
Some phenomenological consequences are also discussed.
%
\end{abstract}

\vskip 1in
\centerline{PACS numbers: 11.30.Er, 14.80.Er \hfill}
%
\newpage
%
\pagestyle{plain}

More than thirty years after its experimental discovery, the origin of
CP violation still remains very much a mystery.  In the
widely accepted Standard Model and many other models,
CP violation is a result of the complex parameters\cite{ref:km} allowed
in the Lagrangian.  For many physicists, such mundane explanation of
the origin of the violation of CP symmetry is not very satisfactory.
In an effort to understand it at a deeper level,
many different schemes have been conceived in the literature.
A popular alternative is to require CP symmetry at the Lagrangian level and
allow its nonconservation only in the vacuum.  Such scheme are commonly
termed spontaneous CP violation\cite{ref:tdlee}.
Another even more ambitious
attempt is to consider CP as an exact symmetry at the tree level but allow
its nonconservation only when the quantum effects are included.
To realize such scheme within the perturbative framework, one naturally
requires the CP violation to be spontaneous in origin also.
Therefore, the model would have a Higgs potential in which its tree level
ground states include a CP conserving one but when the radiative
corrections are included, a CP violating ground state is
selected\cite{ref:rcpx}.
In that case, one can genuinely call the CP violation a quantum mechanical
effect.
It is the aim of this paper to look for a
realistic model of such type.  Through out this paper, the CP symmetry
is assumed to be an exact symmetry of the Lagrangian.

To implement this mechanism in any realistic model, there are two main
obstacles.  The first one is the Georgi-Pais theorem\cite{ref:gp1}.
The theorem assumes
that no fine-tuning of any kind is allowed.  Under such assumption, the
first conclusion one can make is that radiative CP violation is possible
only if the degeneracy of the ground states of the tree level Higgs potential
is such that CP symmetry cannot be asserted.  That is, the
CP violating ground states and the CP conserving ones are degenerate.
Furthermore, Georgi and Pais also proved that radiative breaking can occur
only if the tree level spectrum of the Higgs bosons contain a massless
particle.  Such boson may eventually pick up masses when the radiative
effects are included.  However, under the assumption of no fine-tuning  such
boson is necessarily light.  Since the experimental limit on such
light boson is very strong, it seems very difficult to find a realistic
model under such scheme.

After the result of Georgi and Pais, there are many attempts to get around
the constraint from the theorem.  One can try to go beyond the perturbative
framework \cite{ref:oli} which is beyond our present scope.
Alternatively, one
can relax the no-fine-tuning constraint and permit some fine-tuning
as long as it is technically natural.  (By technically natural, we mean,
in this paper, a set of parameters can be assumed to be much smaller than
the rest of the parameters as long as all the radiative corrections to
these small parameters naturally contains powers of their small tree
level values.)  However, even if the Georgi-Pais theorem is circumvented
by technically natural fine-tuning, its physical origin can still
present itself in the form of the existence of a light Higgs boson in
such model.

An example of such situation appears in the model proposed
by Maekawa\cite{ref:maekawa}.
In the minimal supersymmetric model, it is well-known that spontaneous CP
violation cannot happen at tree level.
However, Maekawa showed that, if
some parameters in the Higgs potential are much smaller than the gauge
coupling, it is possible to have spontaneous CP violation when one-loop
effect is taken into account.
Following Maekawa's, Pomarol\cite{ref:pomarol}
pointed out that the Higgs boson spectrum of
such model contains a light boson whose mass lies in the range that has
already been ruled out by the LEP data\cite{ref:pseudo}.
In general the experimental bound on a pseudoscalar boson is not very strong
because a pseudoscalar boson does not couple linearly to the $Z$ boson
directly.
However, in the minimal supersymmetric model, this bound becomes more
severe as the pseudoscalar boson mass can be related to the scalar boson
mass.

Here we wish to present a simple Peccei-Quinn-type extension\cite{ref:pq}
of Standard Model in which the tree level vacuum is automatically CP
symmetric and the radiative corrections induced by some of the Yukawa
couplings can produce a CP violating vacuum.
The idea here is {\it not} to champion a particular model, but
to show that the fundamental mechanism underlying Ref.\cite{ref:maekawa}
has nothing to do with supersymmetry,
and the problem\cite{ref:pomarol} of a very light Higgs boson facing the
model in Ref.\cite{ref:maekawa} is also
not intrinsic to the mechanism itself.

As in Peccei-Quinn\cite{ref:pq} or supersymmetric\cite{ref:susycpx}
models, we start with two Higgs-boson
doublets.  In general, the following non-hermitian terms
would appear in the Higgs potential,
\begin{equation}
-{\cal L}_{\hbox{\small Higgs}}=
- m_{12}^2 \phi_1^{\dagger} \phi_2
+\lambda_5 (\phi_1^{\dagger} \phi_2)^2
+\lambda_6 \phi_1^{\dagger} \phi_1 \phi_1^{\dagger} \phi_2
+\lambda_7 \phi_1^{\dagger} \phi_2 \phi_2^{\dagger} \phi_2 +\hbox{H.c.}
\end{equation}
Then we impose a Peccei-Quinn type symmetry, $Q_1$, to eliminate
the non-hermitian quartic terms of dimension-4.
However we shall allow the soft terms such as
$- m_{12}^2 \phi_1^{\dagger} \phi_2$ to break the $Q_1$ symmetry.
Beyond tree level, the $\lambda_5$ term as well as other
non-hermitian quartic Higgs couplings,
$\lambda_6$ and $\lambda_7$
will be induced as quantum corrections.

At tree level, since the only non-hermitian coupling is $m_{12}^2$,
the ground state is CP symmetric automatically as the relative
phase between $\langle\phi_1\rangle$ and
$\langle\phi_2\rangle$ is zero.
One may think that, with the induction of
non-hermitian quartic terms from the Higgs-boson loops,
it might be possible to produce a CP
violating ground state by fine-tuning.  However, that is not the case
because the induced quartic non-hermitian couplings will be proportional
to $m_{12}^2$ and cannot be used to balance the tree level coupling,
$m_{12}^2$, no matter how much one tunes.   Even worse, the sign of the
leading contribution to $\lambda_5$ is negative. It was shown in
Ref.\cite{ref:maekawa} that to get a CP violating ground state in the two Higgs
doublet model it is necessary that the induced $\lambda_5$ term is positive.

To induce a positive $\lambda_5$ for our purpose, we need to enlarge the
particle content further.  Here we choose
to enlarge the leptonic sector by two additional right-handed neutrino,
$N_{1R}$ and $N_{2R}$, in addition to the usual lepton doublet $L$.
The spectrum of the model now looks like:
		\begin{equation}
\begin{tabular}{||c|c|c|c||c|c||} \hline
         & $\phi_1$ &  $\phi_2$ & $L$ & $N_1$  & $N_2$ \\ \hline
   $Y$   &  1       &  1         & 1  &   0   &  0   \\
   $Q_1$ &  2       &  0         & 1   & $-1$ &  1   \\
   $Q_2$ &  0       &  0         &  1  &  1   &  1   \\ \hline
\end{tabular} \;,
		\end{equation}
where $Y$ is the hypercharge and $Q_2$ is lepton number symmetry,
which is automatic as far as the dimension-4 couplings are concern.
The relevant Yukawa interactions are
\begin{equation}
-{\cal L}_Y(N) = f_1 \bar{L}N_{1R}\phi_1 + f_2 \bar{L}N_{2R}\phi_2
+\hbox{H.c.}
\end{equation}

We assume the global $Q_1$ symmetry on the hard (dimension-4) terms
but we allows the soft terms to break the symmetry.  They contain, in
addition to the $m_{12}$ term, the following three Majorana mass terms
\begin{equation}
-{\cal L}_{SB}(\hbox{dim-3})=
\mu_{12}N_1^T C N_2 + \mu_{11}N_1^T C N_1 + \mu_{22}N_2^T C N_2
+\hbox{H.c.}
\end{equation}

It is important to note that a discrete symmetry,
\begin{equation}
Z_2: \,\,\, N_1 \rightarrow - N_1, \,\,\, \phi_1 \rightarrow - \phi_1,
\end{equation}
is respected by all terms in the Lagrangian except by
the terms $m^2_{12}$ and $\mu_{12}$.  Therefore it is natural
to fine-tune these two couplings $m^2_{12}$ and $\mu_{12}$
to be small.  Also, one should keep
in mind that the Majorana masses break the lepton number symmetry $Q_2$
softly.  Note that $\mu_{11}, \mu_{22}$ mass terms are $SU(2)\times U(1)$
invariant.  Therefore, their values can in principle be
much larger than the $SU(2)$ breaking scale.

Before we get into the discussion of CP symmetry breaking, it is also
interesting to note that if one set the coupling $m_{12}$ to zero, it will
get divergent contribution induced by the $\mu_{12}$ term, however the
divergence is only logarithmic with coefficient proportional to
$\mu_{12}(\mu_{11} + \mu_{22})f_1 f_2$.
In addition, since the couplings $\lambda_6$ and
$\lambda_7$ are also forbidden by the $Z_2$ discrete symmetry, their
induced values will be proportional to $\mu_{12}$ or $m_{12}$ also.
Therefore, by fine-tuning the parameters $\mu_{12}$ and $m_{12}$ to be
small one can make all the couplings which are forbidden by the $Z_2$
symmetry small (relative to the dimensional parameters
$\mu_{11}$ and $\mu_{22}$). Near this limit, one can find a CP violating
ground state.

To break CP symmetry spontaneously, the loop-induced $\lambda_5$
must have a positive sign.  This can be achieved by diagrams in Fig.~1.
\begin{equation}
\lambda_5=-{f_{11}^2f_{22}^2\over 16 \pi^2}
{\mu_{11}\mu_{22} \over \mu_{11}^2-\mu_{22}^2}
\log {\mu_{11}^2\over \mu_{22}^2}   \ .
\label{eq:l5}
\end{equation}
The positive sign of $\lambda_5$ can always be achieved when $\mu_{11}$
and $\mu_{22}$ are of opposite signs.

The minimization of the potential for the most general couplings has
already been done in Ref.\cite{ref:maekawa}.
Parametrizing the vacuum expectation
value (VEV) as $\langle\phi_1\rangle=v_1 e^{i\delta}/\sqrt{2}$ and
$\langle\phi_2\rangle=v_2/\sqrt{2}$, we obtain
\begin{equation}
\cos\,\delta = {2m_{12}^2 - \lambda_6 v_1^2 - \lambda_7 v_2^2 \over
4\lambda_5 v_1 v_2}.
\end{equation}
(The condition that the vacuum preserves $U(1)_{EM}$ is also analyzed
in Ref.\cite{ref:maekawa}).
First of all, since $\lambda_6 v_1^2$ and $\lambda_7 v_2^2$ are
simultaneously one-loop induced and $Z_2$ breaking, they are naturally
small compared with $m_{12}^2$, which is a tree level $Z_2$ breaking term,
and therefore negligible.  To have a significant CP violating phase
$\delta$ we need $m_{12}^2$ to be of the same order as
$\lambda_5 v_1 v_2 $.
This requires the
fine-tuning of the tree level coupling $m_{12}$.  The fine-tuning is
technically natural if we simultaneously fine-tune both $m_{12}$ and
$\mu_{12}$ because they are the only two tree-level terms
forbidden by the $Z_2$ symmetry.  Therefore, one has arrived at a model
in which the CP symmetry is spontaneously broken by the radiative
correction.  (In contrast, it is easy to show that the supersymmetric
models such as the one proposed by Maekawa\cite{ref:maekawa}
is not technically natural).

Since we have chosen to extend the lepton sector of the Standard Model,
we shall next discuss the structure of the neutrino masses in the model.
Consider the one-generation case.  The Majorana mass matrix of the model
can be written as
		\begin{equation}
\begin{tabular}{||c|c|c|c||}
\hline         & $\nu^C$    & $N_{1R}$     & $N_{2R}$ \\ \hline
      $\nu^C$  &  0         & $f_1{v_1\over 2\sqrt{2}}$
               &  $f_2{v_2\over 2\sqrt{2}}$    \\
      $N_{1R}$ &  $f_1{v_1\over 2\sqrt{2}}$  & $\mu_{11}$
               &  ${1\over 2}\mu_{12}$    \\
      $N_{2R}$ &  $f_2{v_2\over 2\sqrt{2}}$  & ${1\over 2}\mu_{12}$
               &  $\mu_{22}$    \\ \hline
\end{tabular} \;.
		\end{equation}
where $\nu^C$ is from the usual left-handed neutrino.
The neutrino spectrum is
nothing but the usually see-saw spectrum of one very light and two very
heavy Majorana particles.  This is especially true if one assumes that
the singlet masses, $|\mu_{11}|$, $|\mu_{22}|$, are much larger than $SU(2)$
breaking scale  (while $|\mu_{12}|$, $\ll |\mu_{11}|, |\mu_{22}|$, is
fine-tuned  be small).  Increase the number of generation by adding
an index to $\nu^C$ is going to simply increase the number of light Majorana
particles.

Next, we deal with the problem of a potentially light pseudoscalar
boson $A$ with a mass $m_A=\sqrt{2\lambda_5} v \sin\delta$ in our model
where $v = \sqrt{v_1^2 + v_2^2}$.
The value of $\lambda_5^{1/2}$ in Eq.~(\ref{eq:l5}) can be naturally
as large as $0.1$.  $m_A$ is easily around 30 GeV.
The masses of the other scalar bosons are usually much larger.
The potential limit on the mass of a pseudoscalar Higgs boson comes  from
LEP experiments. However, in all the analyses\cite{ref:pseudo}, the
pseudoscalar bosons are assumed to be produced by the decay of a
scalar boson $H$.  For the case when the scalar boson is very heavy
(such as $m_H > m_Z$), no limit on $m_A$ has been extracted yet.
One may try to obtain a limit on the pseudoscalar boson by considering the
emission $Z \rightarrow Z^* AA \rightarrow l^+l^-AA$ \cite{ref:hn};
however the branching ratio is about $10^{-8}$, too small for the present
LEP data unless the $ZZAA$ gauge vertex is very large for some peculiar
reason which does not happen in this model.
A pseudoscalar Higgs boson lighter than a $b$ quark can be ruled out by
$b \rightarrow s A$.\cite{ref:pseudo2}

Note that in the limit that the Higgs potential has a
custodial $SU(2) \times SU(2)$ symmetry, the pseudoscalar boson mass is the
same as the charged Higgs boson mass at the tree level\cite{ref:pv}.
Of course, in our case, not only the Higgs potential contains a
parameter which does not respect the custodial symmetry.  The CP
violating ground state we obtained also breaks the custodial symmetry.
These breakings can contribute to the $\rho$ parameter at the one
loop level, however the resulting constraint\cite{ref:pv} are not
significant numerically for the model considered here.

Finally, we shall make a short discussion of the CP phenomenology.
The details of the CP phenomenology depend on how the Higgs doublets
are coupled to the quarks\cite{ref:mp,ref:lw,ref:gjn}.
Since we have only touched upon the leptonic sector to produce radiative
CP violation so far, there are some arbitrariness in deciding how the quarks
are coupled.  Basically, these doublets can couple to quarks in two
different ways.  The first way is to couple one of the Higgs doublets
to the up-type quarks $u_R$ and the other one to the down-type quarks $d_R$.
This is the way chosen in Peccei-Quinn mechanism\cite{ref:pq}.
The second way is to couple both types of quarks $u_R$, $d_R$
to one and same doublet.
We shall only discuss the first option here
even though the second option may also be interesting.


The leading mechanism
of CP violation is through the neutral Higgs boson exchange.  Since the
tree level couplings of the neutral Higgs bosons are flavor conserving,
the leading contribution to the CP violating $\epsilon$ parameter in the kaon
system is through the two-loop diagrams\cite{ref:maekawa,ref:gjn}.
The mechanism also tends to give large contribution to the neutron electric
dipole moment, $d_n$.  While it is generally believed that the neutral Higgs
boson exchange along is not enough to account for all the known CP
phenomenology \cite{ref:lw}, there are however some claims in the
literature Ref.\cite{ref:gjn} that, by properly adjusting parameters, it
is possible to produce large enough $\epsilon$ with small enough $d_n$
in some models of neutral Higgs mediated CP violation. We shall not get
deeply into this complicated and detailed phenomenological issue here
because it is not really directly connected to the main issue we wish to
illuminate.
If it is indeed the case that some tree level flavor changing neutral
currents are needed to produce large enough $\epsilon$, it can easily be
accommodated in this model by a small extension of the quark sector such
as adding a vectorial down quark\cite{ref:bk} which appears in
$E_6$-type grand unified theories. It is also known that the CP
violating $\epsilon'$ of the kaon  decay and CP violating parameters in
hyperon decays are both negligible in this type of model. Detail
analysis of this and various other possibilities will be presented
elsewhere.

The strong CP problem in models with soft breaking of Peccei-Quinn
symmetry is discussed in Ref.\cite{ref:cm}.

Note that even if one does not impose CP
symmetry on the Lagrangian the Higgs sector alone is automatically
CP conserving at tree level.  Therefore, even in models with other
source of CP violation (such as the Kobayashi-Maskawa mechanism\cite{ref:km}),
quantum effects can produce a new independent source of the
CP violation.  Of course, in that case, one can no longer claim that the
CP violation is a quantum mechanical effect.

Finally let us address on the natural scale for the singlet.
Taking the simplifying assumption that
$v_1 \sim v_2 \sim v$, $f_1 \sim f_2 \sim f$ and
$\mu_{11} \sim \mu_{22} \sim \mu$, we can correlate
the pseudoscalar mass $m_A \sim  f^2 v /(4\pi)$ and
the light neutrino mass $m_{\nu} \sim f^2 v^2/\mu$
in  the relation $\mu \sim 4\pi v (m_A/m_\nu)$.
For $m_{\nu} \sim 10$ eV, one needs
$\mu \sim 10^{13}$  GeV.  In a grand unified theory, the singlet scale
$\mu$ is presumably related to the grand unified scale or an intermediate
scale.  Therefore having a high singlet scale is not a
serious problem.

To conclude, we have shown that if one allows fine-tuning which is
technically natural, it is relatively easy to construct models in
which the tree level vacuum is CP invariant while the loop-corrected
potential produces CP nonconserving vacuum.  The basic
ingredient is to impose enough symmetry (the global $Q_1$ symmetry
in our example) on the higher dimensional
terms such that the tree level potential has only one
soft, non-hermitian, symmetry breaking  term.  Then the loop-induced
higher order term can produce the desired CP nonconserving
vacuum through fine-tuning.  To make the fine-tuning technically
natural, one then has to find a smaller symmetry ($Z_2$ in our case)
which can forbid some of
the soft terms while allowing the others.
Since the softer terms typically have smaller discrete symmetry then
the hard terms, such symmetry is not too hard to find either in
general.  The example we provided in this paper is not only simpler than
the supersymmetric models in the literature, it is also more appealing
because the necessary fine-tuning in our case is technically natural.


We wish to thank the HEP Theory group at Argonne National Laboratory
for the hospitality and support while this work is conducted.  We
also wish to thank Alex Pomarol,
Chuan-Hong Chen and Goran Senjanovic for valuable
discussions.  This work is partially supported by Research Corporation and
U.S. Department of Energy, and by National Science Council of Taiwan.
\newpage
\renewcommand\baselinestretch{1}

\newpage
\topmargin=-0.5in
\headheight=0in
\headsep=0in
\textheight=9.7in 		\textwidth=7in
\footheight=2ex 		\footskip=3ex
\oddsidemargin=-.25in		\evensidemargin=-.25in
\hsize=7in
\parskip=0pt
\lineskip=0pt
\abovedisplayskip=3mm plus.3em minus.5em
\belowdisplayskip=3mm plus.3em minus.5em
\abovedisplayshortskip=2mm plus.2em minus.4em
\belowdisplayshortskip=2mm plus.2em minus.4em
		\setlength{\unitlength}{.9mm}
\tolerance=10000

	\begin{figure}
\large
\begin{center}
\begin{picture}(150,150)(0,0)
\thicklines
%
%
\put(50,50){\line(1,0) {50}}
\put(50,50){\line(0,1) {50}}
\put(50,100){\line(1,0) {50}}
\put(100,50){\line(0,1) {50}}
%
%
\multiput(50,50)(-5,-5){6}{\line(-1,-1) {4}}
\multiput(100,50)(5,-5){6}{\line(1,-1) {4}}
\multiput(100,100)(5,5){6}{\line(1,1) {4}}
\multiput(50,100)(-5,5){6}{\line(-1,1) {4}}
%
%
\put(75,50){\vector(1,0){1}}
\put(75,100){\vector(1,0){1}}
\put(50,65){\vector(0,-1){1}}
\put(50,85){\vector(0,1){1}}
\put(100,65){\vector(0,1){1}}
\put(100,85){\vector(0,-1){1}}
%
%
\put(40,110){\vector(1,-1){1}}
\put(40,40){\vector(1,1){1}}
\put(110,110){\vector(1,1){1}}
\put(110,40){\vector(1,-1){1}}
%
%
\put(47,72){\line(1,1){6}}
\put(47,78){\line(1,-1){6}}
\put(97,72){\line(1,1){6}}
\put(97,78){\line(1,-1){6}}
\put(60,75){\makebox(0,0){$\mu_{11}$}}
\put(90,75){\makebox(0,0){$\mu_{22}$}}
%
%
\put(45,85){\makebox(0,0){$N_1$}}
\put(45,65){\makebox(0,0){$N_1$}}
\put(105,85){\makebox(0,0){$N_2$}}
\put(105,65){\makebox(0,0){$N_2$}}
\put(75,45){\makebox(0,0){$L$}}
\put(75,105){\makebox(0,0){$L$}}
\put(30,130){\makebox(0,0){$\phi_1$}}
\put(120,130){\makebox(0,0){$\phi_2^\dagger$}}
\put(30,40){\makebox(0,0){$\phi_1$}}
\put(120,40){\makebox(0,0){$\phi_2^\dagger$}}
\end{picture}
\end{center}
        \end{figure}
\vfill
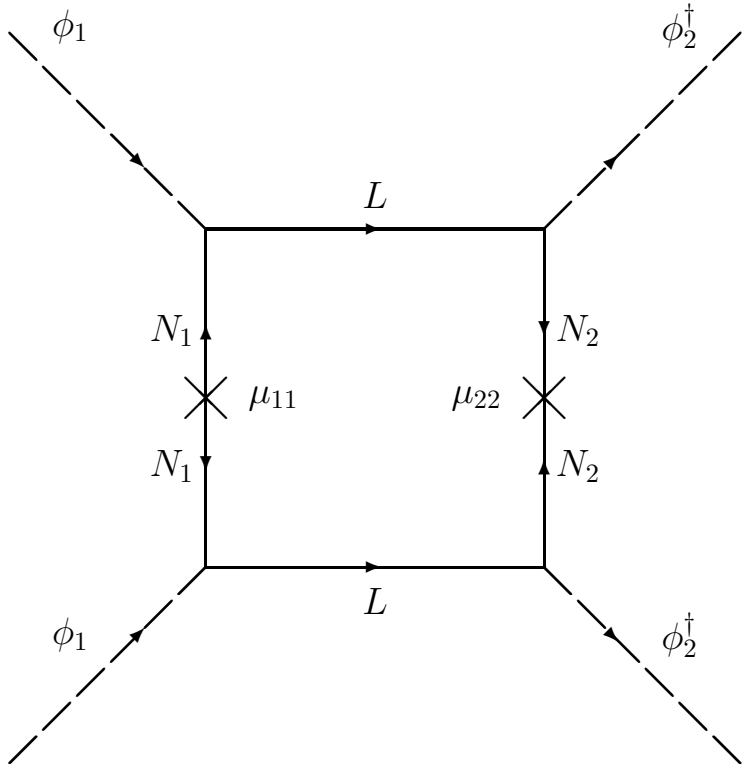
\centerline{Fig.~1 Feynman diagram for the induced vertex
$\lambda_5 (\phi_1\phi_2^\dagger)^2$ via the fermion loop.}
\end{document}